\documentclass[aps,prd,final,twocolumn,letterpaper, superscriptaddress]{revtex4}

\usepackage{amsmath, amssymb, amsthm, cancel, dsfont, enumerate, epstopdf, eucal, fancyhdr, feynmf, gensymb, graphicx, lipsum, mathtools, physics, tikz, verbatim}
\usepackage[caption=false]{subfig}

\usepackage[implicit=true]{hyperref}
\hypersetup{
    pdfnewwindow=true,     
    colorlinks=true,     
    linkcolor=blue,          
    citecolor=blue,      
    filecolor=blue,   
    urlcolor=blue    
}

\begin{document}

\title{AGN variability in the age of VRO}
\author{Cyril Creque-Sarbinowski}
\email{creque@jhu.edu}
\author{Marc Kamionkowski}
\email{mkamion1@jhu.edu}
\author{Bei Zhou}
\email{beizhou@jhu.edu}
\affiliation{Department of Physics and Astronomy, Johns Hopkins University, 3400 N. Charles St., Baltimore, MD 21218, USA}
\date{\today} 

\begin{abstract}
Over the next ten years, the Vera C.\ Rubin Observatory (VRO) will observe $\sim$10 million active galactic nuclei (AGN) with a regular and high cadence.  During this time, the intensities of most of these AGN will fluctuate stochastically. Here, we explore the prospects to quantify precisely these fluctuations with VRO measurements of AGN light curves.  To do so, we suppose that each light curve is described by a damped random walk with a given fluctuation amplitude and correlation time. Theoretical arguments and some current measurements suggest that the correlation timescale and fluctuation amplitude for each AGN may be correlated with other observables. We use an expected-information analysis to calculate the precision with which these parameters will be inferred from the measured light curves.  We find that the measurements will be so precise as to allow the AGN to be separated into up to $\sim 10$ different correlation-timescale bins.  We then show that if the correlation time varies as some power of the luminosity, the normalization and power-law index of that relation will be determined to $\mathcal{O}(10^{-4}\%)$. These results suggest that with VRO, precisely measured variability parameters will take their place alongside spectroscopy in the detailed characterization of individual AGN and in the study of AGN population statistics.  Analogous analyses will be enabled by other time-domain projects, such as CMB-S4.
\end{abstract}

\maketitle

\pagestyle{myheadings}
\markboth{Cyril Creque-Sarbinowski}{AGN variability in the age of VRO}
\thispagestyle{empty}

\section{Introduction}
The intensity of most active galactic nuclei (AGN) is observed to vary on timescales from minutes to decades and in frequency bands ranging from radio to gamma ray~\cite{0907.1415, 1605.09331, 1707.07134, 1802.05717}. X/Gamma-ray variability is thought to arise from the innermost part of the AGN~\cite{1993MNRAS.265..664G, astro-ph/0206190, 0905.4842}, while optical/UV variability from the outer accretion disk as the result of instabilities or X-ray reprocessing~\cite{2001ApJ...555..775C, astro-ph/0409254, astro-ph/0512394}. However, there can be additional contributions in the optical/UV band from the broad-line region and dust torus either due to intrinsic variability~\cite{astro-ph/0301216, 0905.1981} or X-ray reprocessing~\cite{1407.6361, 1502.01977}. In addition, radio variability can also be intrinsic, and if so comes from the inner core of the AGN~\cite{0907.1489}. The variability from higher frequency light can occur on timescales between minutes and years~\cite{1704.08148, 1909.04227, 2001.01105}, while lower frequency light tends to occur on the scales of months to years~\cite{1002.3365}. At the higher end of the variability timescale are changing-look AGN, whose intensity can fluctuate over time periods of decades~\cite{1412.2136, 1509.08393, 1711.08122}. Despite their prevalence in AGN physics, the relationship between AGN type and variability timescale, or even the causes of variability have still not been fully characterized. With the advent of the Vera C. Rubin Observatory (VRO), the flux measurements of over 10 million AGN will be made over the course of 10 years, with first light slated for the end of 2022~\cite{0912.0201}. The huge number of sources, coupled with a vast increase in the number of regular high frequency observation epochs, VRO will allow for unparalleled precision in variability analyses.  

The rich and ubiquitous property of AGN variability has been exploited not only to classify~\cite{0909.1326, 1508.04121} but also to identify AGN~\cite{1008.3143, 1312.4957, 1509.05607, 1904.04844, 2011.08860}. However, attempts to connect variability measurements to physical mechanisms have lacked sufficient data, leaving such connections mostly tenuous~\cite{0711.1013, astro-ph/0502112}. More success has been found in modelling the stochastic nature of these processes~\cite{1009.6011}. Such modelling has seen that most AGN exhibit variability that is well described by a damped random walk (DRW), as shown by analyses involving structure functions, autocorrelation functions, and power spectra~\cite{1004.0276, 1112.0679, 1202.3783, 1604.01773, 1611.08248}, although, some examples of non-DRW AGN have been found~\cite{1505.00360, 1803.06436, 1205.4255}. The DRW model has been used to extract the variability timescale through the use of the structure function~\cite{1701.00005}. In addition, numerical investigations have been used to model AGN variability across multiple timescales~\cite{1909.06374, 1607.04299}. Recently,the variability of 67 AGN were characterized to follow a power law with an index measured to one part in ten~\cite{2108.05389}.  

In this paper, we explore the prospects for studying AGN variability with VRO. Previous analyses of AGN variability have been limited due to small AGN sample size or infrequent visit times, both of which will be remedied with the dawn of VRO. For a single AGN, we forecast that VRO will measure both the variability amplitude and timescale up to $10^3\sigma$. With such precise measurements, we then model a power law relationship between an AGN's bolometric luminosity and its variability with index $\beta_b$ in VRO frequency band $b$. We find that this index will be measured possibly to one part in a million. Therefore, these results suggest that the next decade of observations will lead to a wealth of knowledge in AGN variability.  

This paper is structured as follows. In Sec.~\ref{sec:form} we present the formalism for measuring the variability amplitude and timescale in the context of a single AGN and a population of AGN. We then follow up this formalism in Sec.~\ref{sec:4cast} and present estimators in order to quantify variability. Moreover, using these estimators we make an expected information forecast to VRO's sensitivity in measuring both variability parameters for a single AGN, along with their covariance. Then, we present the analogous calculation for the theoretical best sensitivity to measuring a power law relationship between AGN bolometric luminosity and variability. We discuss these results and conclude in Secs.~\ref{sec:disc} and \ref{sec:conc}, respectively. 
\section{Formalism}\label{sec:form}
Assume the intensities from a population of AGN have been measured over time. We pursue a description of the variability of these intensities through the use of the two-point correlation of their intensities. In this vein, we first present the autocorrelation function for a single AGN. Then, for a population of AGN, we extend the presentation of a single AGN and model a relationship between the bolometric luminosity of an AGN with both its variability parameters through a power law. 
\subsection{Single AGN}\label{subsec:sing_agn}
Let $I_b^j(t)$ be the observed intensity of AGN $j$ in frequency band $b$ and $\bar{I}^j_b$ be its time average . With these two quantities, define $\delta_b^j(t) = I^j_b(t)/\bar{I}_b^j - 1$ to be the observed variability of AGN $j$ in frequency band $b$. We describe the statistical properties of the observed variability of AGN $j$ in frequency band $b$ in terms of the observed two-point variability correlation function $\langle \delta_b^j(t_1)\delta_b^j(t_2)\rangle$. If the underlying mechanism creating the observed signals is independent of time for the duration of observation, then the two-point correlation function is homogeneous in time (i.e. stationary) and thus only a function of the time lag $t = |t_2 - t_1|$, $\langle \delta_b^j(t_1) \delta_b^j(t_2) \rangle = \langle \delta_b^j(t' + t)\delta_b^j(t')\rangle$. 

With this assumption, we model the observed two-point variability correlation function for a single AGN as   
\begin{align}\label{eq:var_corr}
\left\langle \delta_b^j(t' + t)\delta_b^j(t')\right\rangle &= \xi_{bb}^j(t) + \Delta t_b\sigma_{jb}^2 \delta(t),\\
\xi_{bb}^j(t) &\equiv A_{jb}^2 e^{-|t|/\bar{t}^o_{jb}},\label{eq:var_cf}
\end{align}
with $\xi_{bb}^j(t)$ the two-point variability correlation function taking the form of a damped random walk~\cite{1004.0276, 1202.3783, 1312.3966}, $A_{jb}$ the variability amplitude and $\bar{t}^o_{jb}$ the variability timescale of AGN $j$ in the observer's frame, $\Delta t_b$ the temporal resolution of the experiment, and $\delta(t)$ the Dirac delta function. Moreover, $\sigma_{jb}^2$ is the variance of a white noise process representing photometric error in an AGN's intensity measurements. We assume this noise does not correlate with any AGN's variability. Note that the observed two-point variability correlation function contains instrumental noise, while the two-point variability correlation function does not. Due to either AGN physics or instrument properties, all quantities mentioned depend on the observing frequency band $b$. Furthermore, due to cosmic redshifting, the variability timescale $\bar{t}^r_{jb}$ of an AGN located at redshift $z$ as measured in its rest frame is related to the observer frame analog through the expression $\bar{t}_{jb}^o = (1 + z)\bar{t}_{jb}^r$. 

\begin{figure}
\includegraphics[width = 0.49\textwidth]{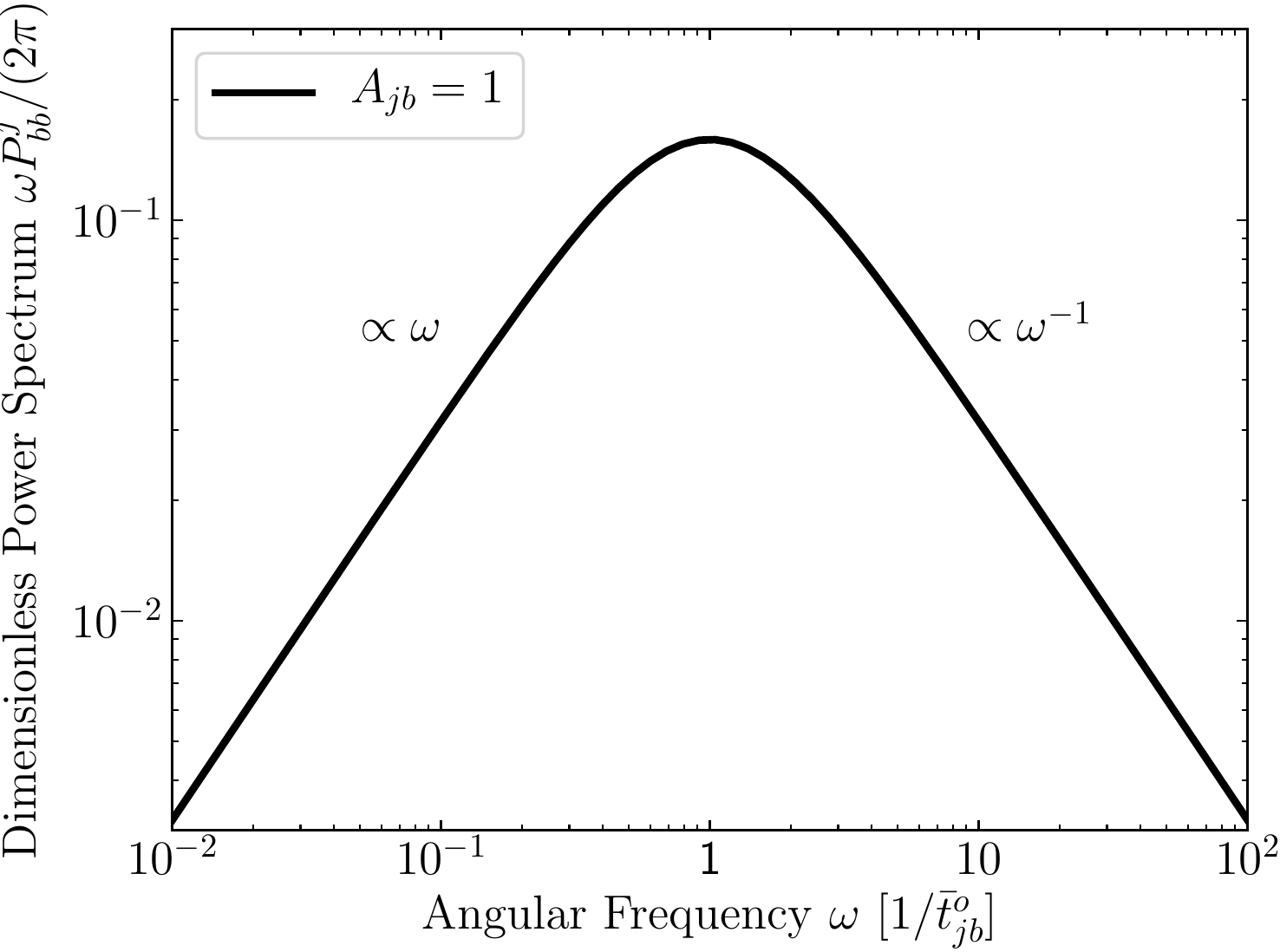}
\caption{The dimensionless variability power spectrum $\omega P_{bb}^j/(2\pi)$ for an AGN with variability amplitude $A_{jb} = 1$ using Eq.~\eqref{eq:var_ps}. The peak of this power spectrum occurs at $\omega = \bar{t}_{jb}^{-1}$, with amplitude $A_{jb}^2/(2\pi)$. For frequencies smaller than this peak it rises as $\omega$, and for frequencies larger it falls as $\omega^{-1}$. Since we plot the angular frequency in units of $\bar{t}_{jb}^{-1}$, its value is arbitrary.}\label{fig:single_PS}
\end{figure}

With this expression of the observed correlation function, its Fourier transform is
\begin{align}
\frac{1}{2\pi}\int d\omega' \left\langle \tilde{\delta}_b^j(\omega)\tilde{\delta}_b^{j*}(\omega')\right\rangle &= P_{bb}^j(\omega) + \Delta t_b\sigma_{jb}^2,\\
P_{bb}^j(\omega) &\equiv \frac{2A_{jb}^2\bar{t}_{jb}}{1 + (\omega \bar{t}_{jb})^2},\label{eq:var_ps}
\end{align}
with $P_{bb}^j(\omega)$ the variability power spectrum for AGN $j$ in frequency band $b$. Here and in what follows, we use Fourier convention $f(t) = (2\pi)^{-1}\int d\omega e^{-i\omega t} \tilde{f}(k)$ and $\tilde{f}(\omega) = \int dt e^{i\omega t}f(t)$. We plot an example variability power spectrum for a single AGN in Fig.~\ref{fig:single_PS}.
 
\subsection{AGN Population}\label{subsec:pop_agn}
In addition to modelling each AGN individually, we also model the two-point variability correlation function of a single AGN from a set of population parameters. Define $\xi_{bb}(t, z, L)$ to be the variability correlation function for an AGN located at redshift $z$ with bolometric luminosity $L$,
\begin{align}
\xi_{bb}(t, z, L) &= A^2_b(L)\exp\left[-|t|/\bar{t}^o_b(z, L)\right].
\end{align} 
In writing this expression, we implicitly assume that AGN variability is only characterized by two parameters: its redshift and bolometric luminosity. In addition to this correlation function, we define  the power spectrum $P_{bb}(\omega, z, L)$ analogously. 

Typically, AGN with higher bolometric luminosities are more massive. Since variability on timescales smaller than the light crossing time of the emitted object is suppressed, the larger an AGN, the larger its expected variability timescale. Thus, we model the relationship between AGN variability and bolometric luminosity as
\begin{align}
A_b(L) &= A_{b},\\
\bar{t}^o_b(z, L) &= \bar{t}_b^r(1 + z)\left(\frac{L}{L_b}\right)^{\beta_b}\label{eq:tLrel},
\end{align}
with $L_b = L_{\rm bol}(m_{\rm lim}^b, z_{\rm min})$ a normalization constant chosen to be the dimmest expected observed AGN. In this expression, we assumed all AGN to have the same variability amplitude for simplicity. Thus, a population of AGN is described by the three parameters $A_b, \bar{t}_b^r,$ and $\beta_b$. Recently, 67 AGN were found to follow a similar variability timescale relation, with the mass of the AGN as the the only dependent parameter, and the index $\beta\sim 0.23$~\cite{2108.05389}.
 
Let $dN_{\rm AGN}/dz dL$ be the redshift and bolometric luminosity distribution of this population of AGN. Then these AGN are distributed throughout the Universe according to
\begin{align}\label{eq:dNAGNdLdz}
\frac{dN_{\rm AGN}(z, L)}{dz dL} &= \frac{dV(z)}{dz}\frac{dn(z, L)}{dL},
\end{align}
with $dn(z, L)/dL$ the AGN luminosity function, $dV(z)/dz= 4\pi f_{\rm sky} r(z)^2 dr(z)/dz$ the comoving volume observed over a fraction $f_{\rm sky}$ of the sky, $r(z) = \int_0^z |dr/dz| dz$ the comoving radial distance to a redshift $z$, $dr/dz = - c/(1 + z)H(z)$ its redshift derivative, $c$ the speed of light, and $H^2(z) = H_0^2\left[\Omega_m(1 + z)^3 + (1 - \Omega_m)\right]^{-1/2}$ the Hubble parameter. We use Planck 2018 $\Lambda$CDM parameters $H_0 = 2.18\times 10^{-18} {\rm s}^{-1}$ and $\Omega_m = 0.315$~\cite{1807.06209}, along with the Full AGN luminosity function in Table 3 from Ref.~\cite{astro-ph/0605678}. 

\begin{figure}
\includegraphics[width = 0.49\textwidth]{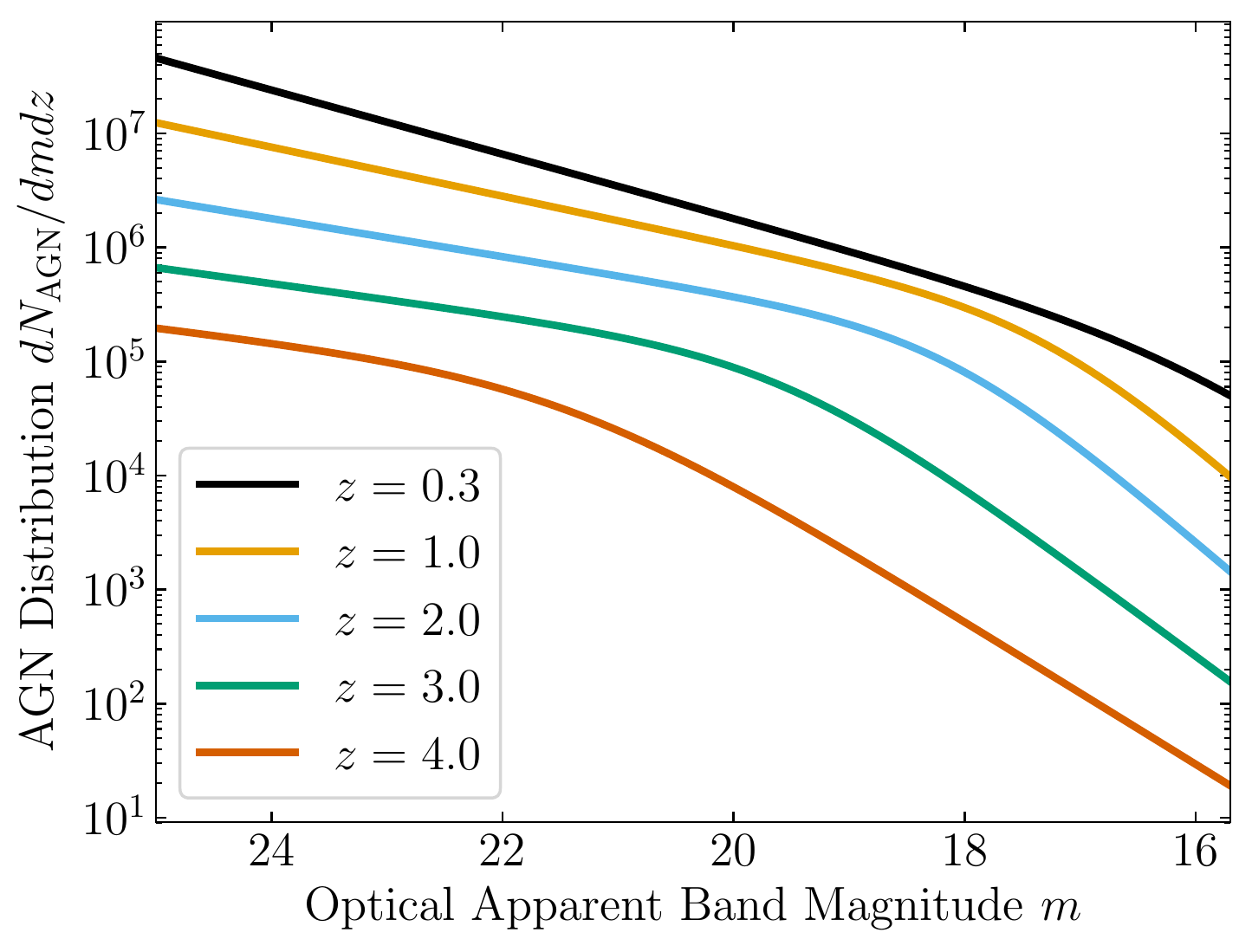}
\caption{The distribution $dN^b_{\rm AGN}/dm_b dz$ of AGN as a function of the apparent magnitude in a band $b$ at redshifts $z\in\{0.3, 1.0, 2.0, 3.0, 4.0\}$ as given by Eq.~\eqref{eq:dNAGNdmdz}. In order to show the full range of this distribution, we do not include the theta function factor.}\label{fig:dNAGNdmdz}
\end{figure}

Given a cosmological distribution of AGN, only those that appear bright enough will be observed. More specifically, given a limiting apparent magnitude $m^{\rm lim}_b$ in a band $b$, the distribution of observed AGN in that band is
\begin{align}\label{eq:dNAGNdmdz}
\frac{dN_{\rm AGN}^b}{dz dm_b} &= \Theta(m^{\rm lim}_b - m_b)\frac{dL_{\rm bol}}{dm_b}\frac{dN_{\rm AGN}[z, L_{\rm bol}(z, m_b)]}{dz dL_{\rm bol}},
\end{align}
with $\Theta(x)$ the Heaviside theta function, $L_{\rm bol}(z, m_b) = K_b(m_b) 4\pi d_L(z)^2 \delta \nu_b F_{\rm AB} 10^{-(2/5)m_b}$ the bolometric luminosity for an AGN with apparent magnitude $m_b$ in frequency band $b$ located at redshift $z$, and $dL_{\rm bol}/dm_b = -(2/5)\log(10)L_{\rm bol}(m_b, z)$ its apparent magnitude derivative. Moreover, $K_b(m_b)$ is the bolometric correction function to convert from the emitted luminosity in band $b$ to bolometric luminosity of the source, $d_L(z) = (1 + z)r(z)$ the luminosity distance, $\delta \nu_b$ the frequency bandwidth of band $b$, and $F_{\rm AB} = 3.631\times 10^{-23}\ {\rm W}{\rm Hz}^{-1}{\rm m}^{-2}$. Note that in this expression for the bolometric intensity, we assume the observed intensity is roughly constant across the entire bandwidth, and that the redshifted frequency does not alter the intensity in each band significantly. In general, the bolometric correction within the optical range is a function of the bolometric luminosity of the source. However, across all bolometric luminosities the correction changes only up to $20\%$, and inversion of this expression can only be done numerically. Thus, for simplicity, we adopt that $K_b(m_b) = 10$ for all magnitudes and bands~\cite{astro-ph/0605678}. We plot the observed AGN distribution, without the theta function, in Fig.~\ref{fig:dNAGNdmdz}. 
\begin{comment}
with $N^b_{\rm AGN}$ the total number of AGN observed  in frequency band $b$. The redshift integral is taken over the observed redshift ranges for the AGN population, while the apparent magnitude integral from zero to infinity. In practice, there are no observed AGN with zero apparent magnitude, and a nonzero cutoff $m_{\rm cut}^b = 15.7$ is imposed across all bands. 
\end{comment}

\section{Forecasts}\label{sec:4cast}
One may measure both the variability amplitude and timescale of an AGN from measuring only its variabiliy correlation function at different lag times. However, the relation between the observed correlation function at any lag time and the true underlying stochastic process becomes increasingly inaccurate for variability timescales smaller than the cadence and larger than the observational period. On the other hand, the observed power spectrum is accurate for all Fourier modes well within these limits.

\begin{figure}[ht]
\renewcommand{\thesubfigure}{c}
\subfloat[]{\includegraphics[width = \linewidth]{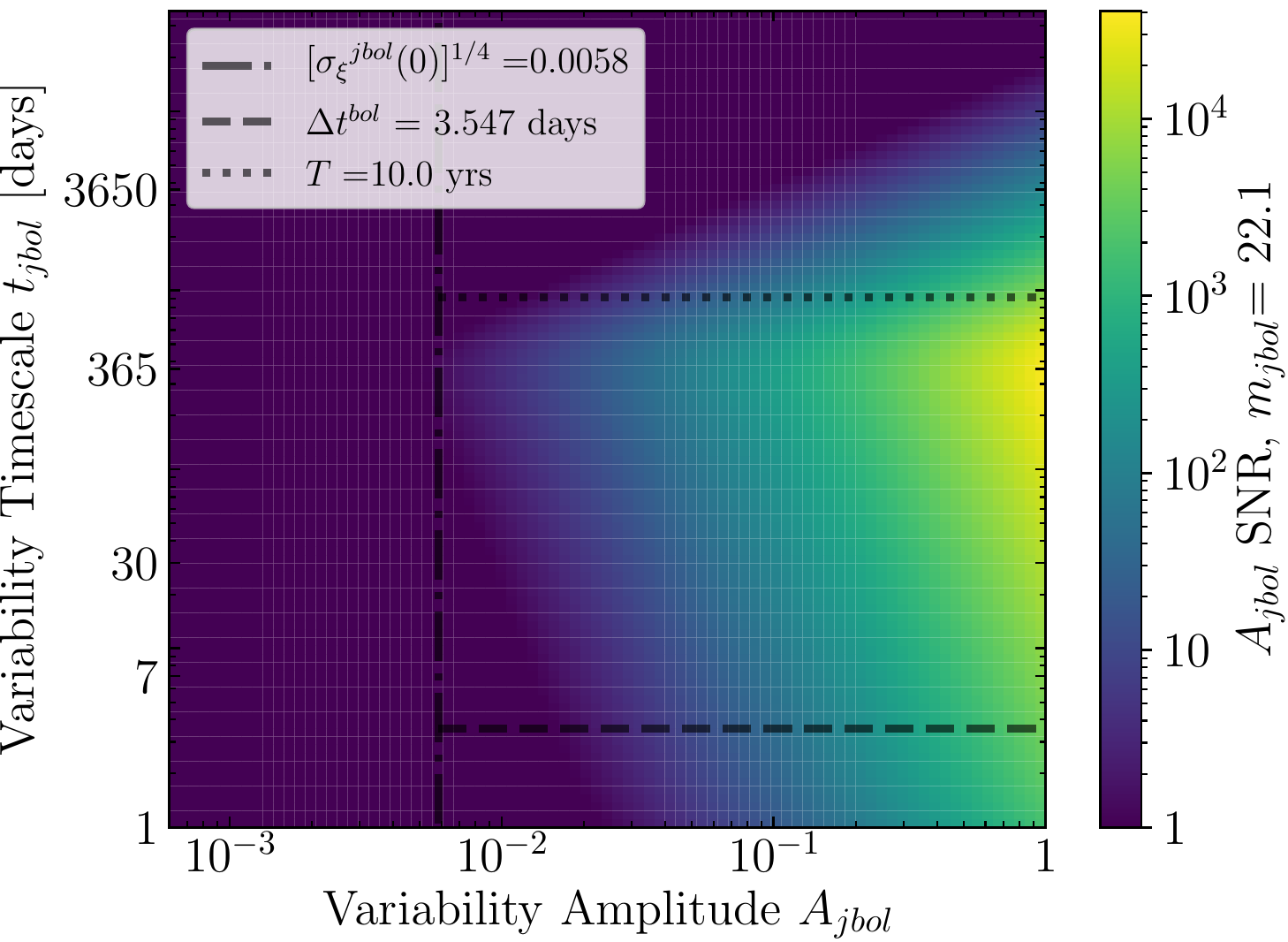}}
\caption{The signal-to-noise-ratio (SNR) in measuring the variability amplitude $A_{jb}$ of AGN $j$ in VRO frequency band $b$ as a function of the measured variability amplitude and timescale. This forecast is made using the dimmest AGN to be observed $\bar{I}^j_b = I_{\rm bol}(m_{\rm lim}^b, z)$ with noise $\sigma_{jb}$, temporal resolution $\Delta t_b$, and observation run $T = 10$  years. The three bands presented are VRO's (a) $u$ (b) $i$ and (c) inferred bolometric bands. When the variability amplitude drops below the noise threshold, as indicated by the dot-dash line, the error becomes too large and the measurement fidelity significantly drops. If the variability timescale is larger than the observation time, as indicated by the dotted line, then all intensity measurements are maximally correlated and the SNR saturates to a constant signal. On the other hand, if the variability timescale is smaller than the temporal resolution, as shown by the dashed line, then each measurement is maximally independent and thus the SNR saturates once more. }\label{fig:ASNR}
\end{figure}

\begin{figure*}[t]
\includegraphics[width = 0.8\textwidth]{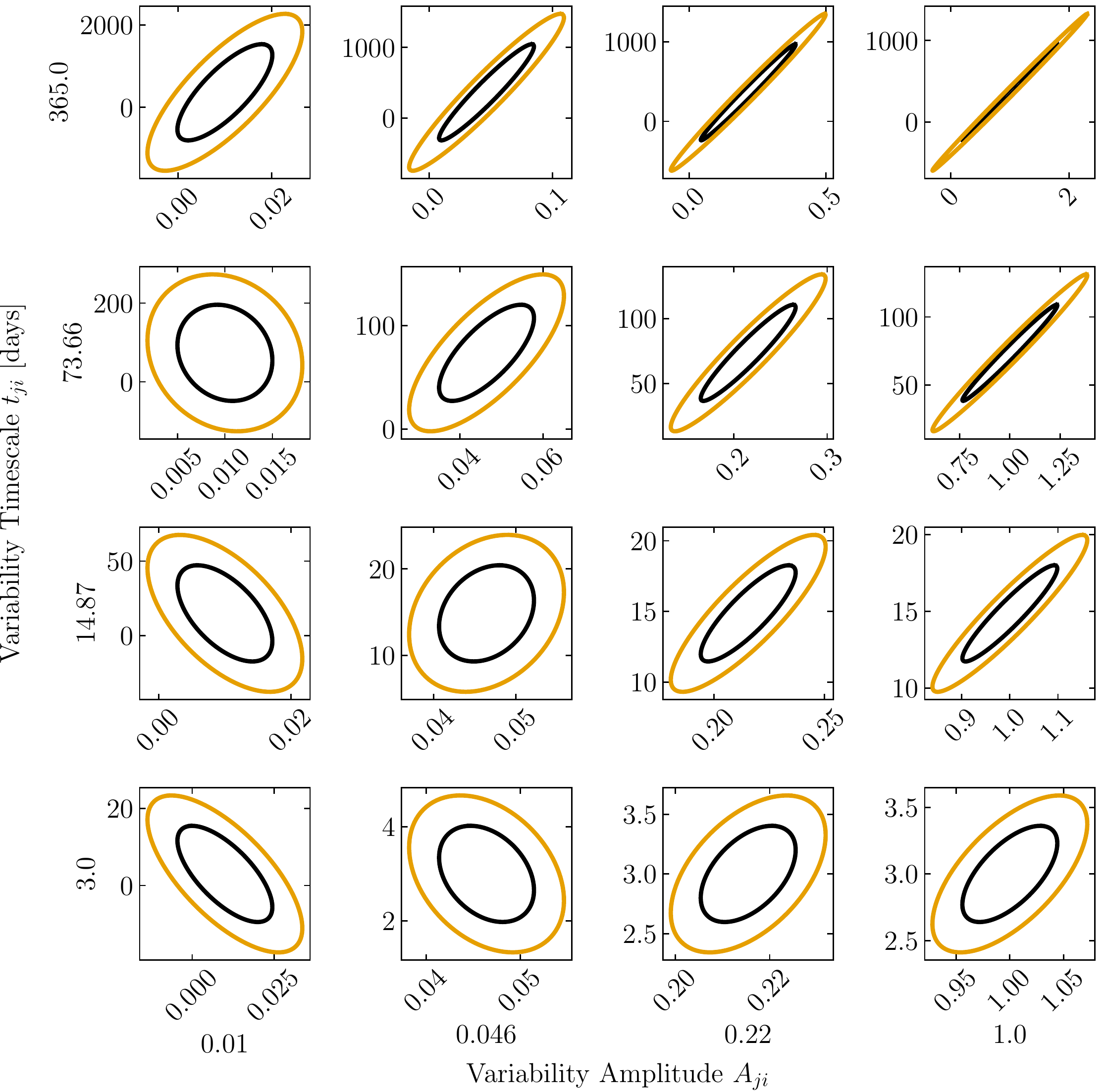}
\caption{The covariance between the variability amplitude $A_{ji}$ and observed timescale $\bar{t}^o_{ji}$ in VRO's $i$ band for an AGN with apparent magnitude $m_i = 22.3$, corresponding to the average AGN apparent magnitude in the $i$ band. We show the covariance assuming fiducial parameters $A_{ji} \in \{10^{-2}, 4.6\times10^{-2}, 2.2\times 10^{-1}, 1 \}$ and $\bar{t}^o_{ji} = \{3\ {\rm days}, 14.87\ {\rm days}, 73.66\ {\rm days}, 365\ {\rm days}\}$. The black circles indicates $1\sigma\ (68\%)$ confidence, and the yellow $2\sigma\ (95\%)$. We note that these results hold for most AGN magnitudes and VRO frequency bands, as the AGN included in this analysis are all assumed to be detected at high signal to noise. For low amplitude and variability timescale, only modes in the white noise regime, $P \propto (A^2 \bar{t})\omega^0$, of the power spectrum are probed, and so there is negative correlation between the two parameters. As the two parameters increase, the red noise regime , $P \propto (A^2/\bar{t})\omega^{-2}$, leads to a positive correlation.}\label{fig:ellipse}
\end{figure*}

\begin{figure*}[ht]
\includegraphics[width = \textwidth]{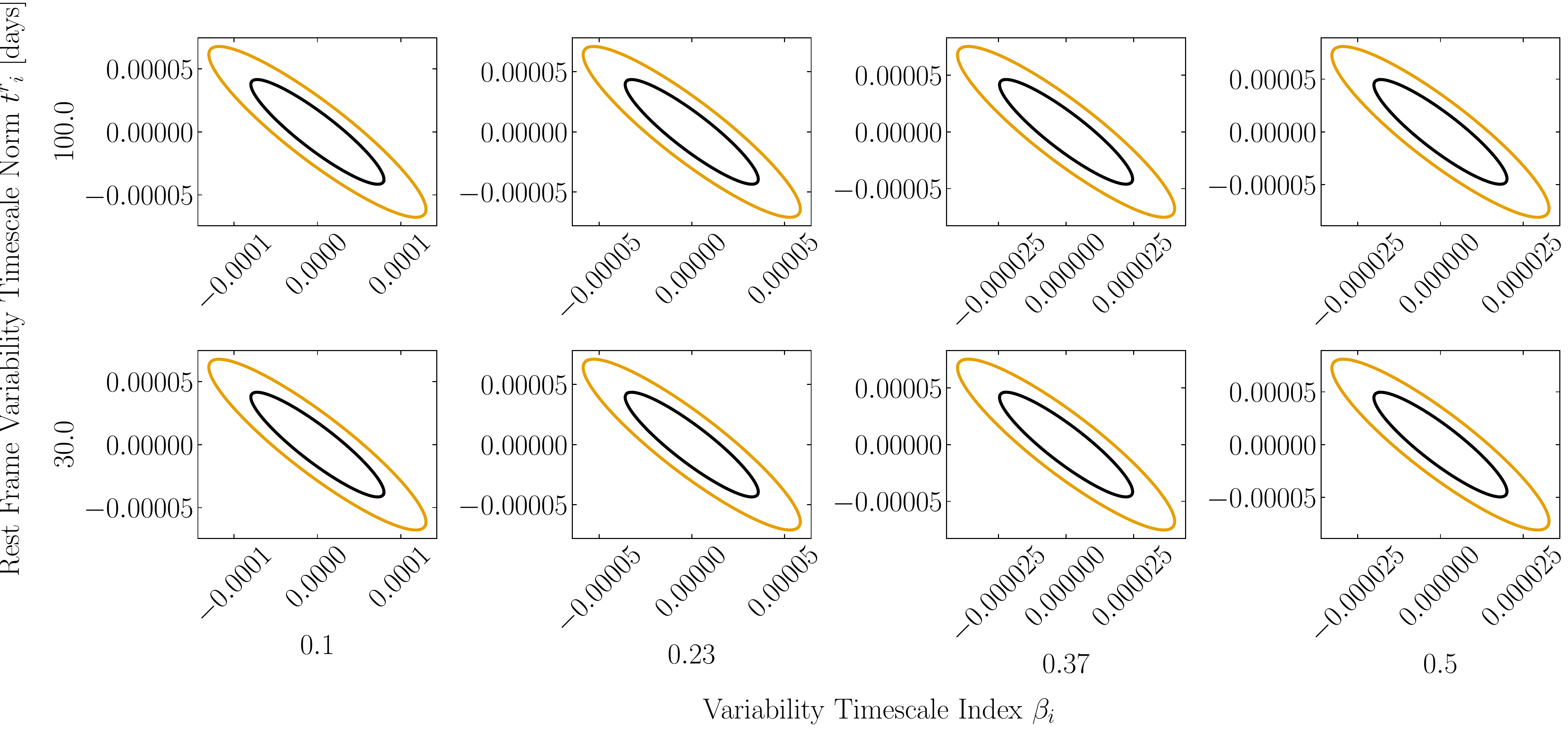}
\caption{The covariance between the variability timescale index and norm in VRO's $i$ band in terms of the fractional differences $(\beta - \beta_{\rm fid})/\beta_{\rm fid}$ and $(\bar{t}^r - \bar{t}^r_{\rm fid})/\bar{t}^r_{\rm fid}$. We take fiducial parameters $\beta_i \in \{0.1, 0.23, 0.37, 0.5\}$ and $\bar{t}_i^r = \{30\ {\rm days}, 100\ {\rm days}\}$. The black circles indicate $1\sigma\ (68\%)$ confidence, and the yellow $2\sigma\ (95\%)$. We note that these results hold for all VRO frequency bands, as VRO is limited not by instrumental noise. We take $A_i = 1$. Since increases in both the index $\beta$ and the norm $\bar{t}_r$ increase the observed variability timescale, they are anti-correlated. 
}\label{fig:ellipse2}
\end{figure*}

In this section, we use the expected information from the power spectrum estimators for a single AGN and a population of AGN to forecast VRO's ability to measure various variability parameters. The analysis, discussed below, leads to the signal-to-noise results for the variability amplitude in Fig.~\ref{fig:ASNR}. Furthermore, we show the covariance between the variability amplitude and timescale in Fig.~\ref{fig:ellipse} and the variability index and timescale in Fig.~\ref{fig:ellipse2}. All figures are done for a representative sample of VRO's frequency bands. 

\begin{figure}
\includegraphics[width = 0.49\textwidth]{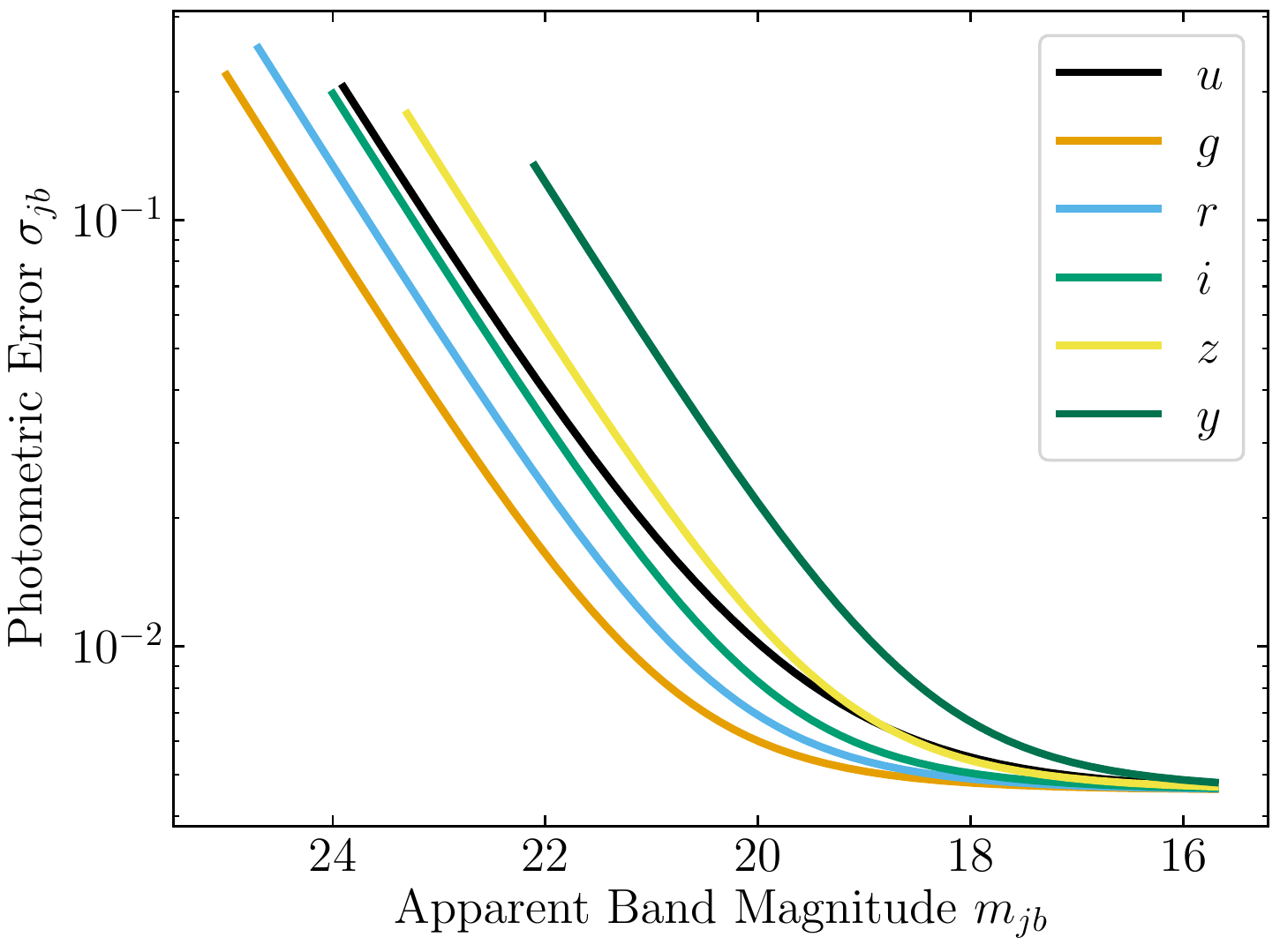}
\caption{The VRO photometric error for an AGN's variability in each VRO frequency band $b \in \{u,g,r,i,z,y\}$, as well as for the bolometric variability error, as given by Eq.\eqref{eq:photo_err} and Eq.~\eqref{eq:bol_err}. The magnitudes plotted ranges from the $5\sigma$ apparent magnitude limit in the corresponding band, shown in Table.~\ref{tab:VRO}, to  the theoretical value for the brightest AGN that will be observed $m_{\rm cut}^b = 15.7$. }\label{fig:photo_err}
\end{figure}

We model the individual band errors as a sum of Poissonian shot noise from source, Poissonian shot noise from the sky, Gaussian instrumental noise, and systematic error. More specifically, for VRO we use the fit given by Ref.~\cite{0912.0201} and propagate the error from apparent magnitude to variability, 
\begin{align}\label{eq:photo_err}
\sigma_{jb}^2 &= \frac{2}{5}\log(10)\left[\sigma_{\rm sys}^2 + \left(\sigma^{\rm rand}_{jb}\right)^2\right],\\
\left(\sigma^{\rm rand}_{jb}\right)^2 &= (0.04 - \gamma_{b})x_{jb} + \gamma_{b}x^{2}_{jb},   
\end{align} 
with $x_{jb} = 10.0^{(2/5)(m_{jb} - m_5^b)}$, $m_{jb}$ the apparent magnitude of AGN $j$, and $m_5^b$ the $5\sigma$ depth for point sources, both defined in frequency band $b$. The fitted parameter $\gamma_b$ depends on sky brightness, readout noise, and other factors. We show the relevant experimental parameters for each band in Table.~\ref{tab:VRO} and plot the photometric error in Fig.~\ref{fig:photo_err}. For all bands we take the limiting apparent magnitude to be the 5$\sigma$ point source depth, $m_{\rm lim}^b = m_5^b$. Moreover, define $n_{\rm vis}^b = T/\Delta t_b + 1$ to be the number of visits to AGN $j$ in frequency band $b$ and $n_{\rm vis} = \sum_b n_{\rm vis}^b$ the total number of visits. Note that we assume that all AGN are visited an equal number of times across all frequency bands. 
\begin{table}[ht]
\begin{tabular}{|c|c|c|c|c|c|c|}
\hline
$b  $ & $u$ & $g$ & $r$ & $i$ & $z$ & $y$\\
\hline
$\gamma_b$ & 0.038 & 0.039 & 0.039 & 0.039 & 0.039 & 0.039\\
\hline
$m_5^b$ & 23.78 & 24.81 & 24.35 & 23.92 & 23.34 & 22.45 \\
\hline
$n_{\rm vis}^b$ & 70 & 100 & 230 & 230 & 200 & 200\\
\hline
$\Delta t_b\ [{\rm days}]$ & 52.90 & 36.87 & 15.94 & 15.94 & 18.34 & 18.34\\
\hline
\end{tabular}\caption{The experimental parameters for VRO.\label{tab:VRO}}
\end{table}

\begin{figure}
\subfloat[]{\includegraphics[width = \linewidth]{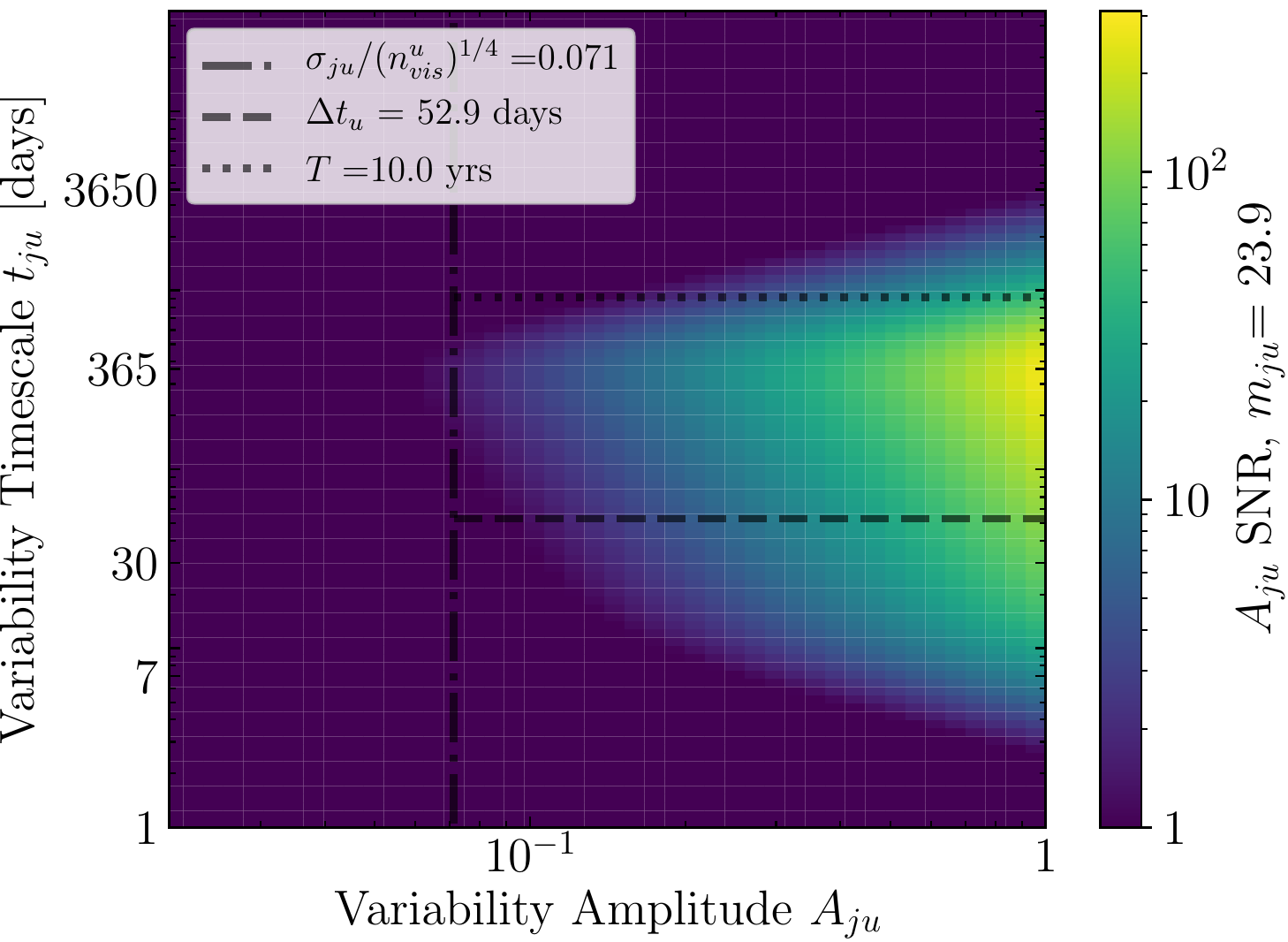}}
\qquad
\subfloat[]{\includegraphics[width = \linewidth]{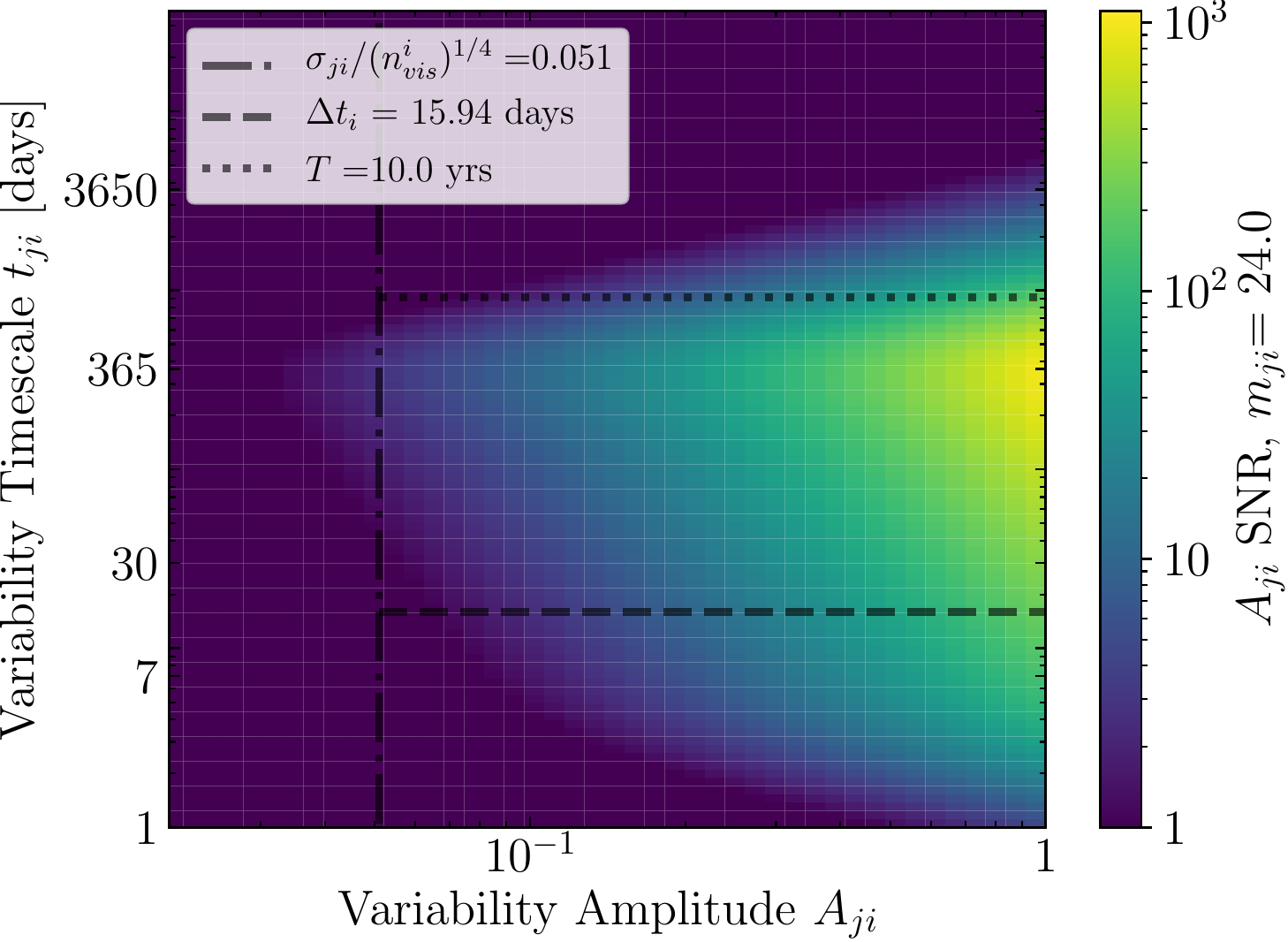}}
\end{figure}
\subsection{Single AGN}
For notational simplicity, we assume all AGN are observed for the same duration $T_j = T$ and sampled at the same times. However, we allow for different sampling between different frequency bands $b$. Under the null hypothesis, AGN undergo no variability and thus the noise for the power spectrum estimator of AGN $j$ in band $b$ is 
\begin{align}
\left[\sigma_{\rm null}^{jb}(\omega)\right]^2 &= 2\left[\Delta t_b \sigma^2_{jb}\right]^2.
\end{align}
We also combine the information of all bands through the use of bolometric corrections. First we calculate the bolometric correlation function as given by a particular band. Since the bolometric intensity is approximately linear in the band intensity, their fractional errors are the same. Then, we inverse-variance weigh each band to obtain an estimate of the actual bolometric correlation function. Therefore, the error $\sigma_{\rm null}^{j {\rm bol}}$ in measuring this correlation function is 
\begin{align}\label{eq:bol_err}
\left[\sigma_{\rm null}^{j{\rm bol}}(\omega)\right]^{-2} &= \sum_b \left[\sigma_{\rm null}^{jb}(\omega)\right]^{-2}. 
\end{align}
Since the bolometric band is a combination of measurements done in different bands with different temporal resolutions - the bolometric band has unequal, but periodic, temporal spacing in measurements. Moreover, not every temporal spacing has an equal number of measurements. Rather than model this spacing, we take an equal-time temporal resolution $\Delta t_{\rm bol} = T/(n_{\rm vis} - 1)$, with the condition that null-hypothesis forecasts using this resolution are upper bounds. 

Moreover, we note that given the bolometric correlation function, we can invert the bolometric corrections in order to translate the bolometric error into the error in any particular band $b$. Thus, through inverse variance weighing, the bolometric band represents the optimal sensitivity for any particular band. 
 
With the null-hypothesis power spectrum noise in hand, we use the expected information matrix to infer the covariance matrix for our AGN parameters. We plot the signal-to-noise of measurements for the variability amplitude for a single AGN under the null hypothesis in Fig.~\ref{fig:ASNR}.

To calculate the covariance between the variability amplitude $A_{jb}$ timescale $\bar{t}_{jb}$ once a signal is detected, we must include the correlations from the signal. Therefore, the noise for the power spectrum $P_{bb}^j(\omega)$ estimator is now 
\begin{align}
\left[\sigma_P^{jb}(\omega)\right]^2 &= 2\left[P_{bb}^j(\omega) + \Delta t_b \sigma^2_{jb}\right]^2.
\end{align}
Under the non-null hypothesis, there is a covariance induced in the Fourier amplitudes inferred between different bands. Therefore, in order to asses the ability of VRO to synthesize information from different bands, we assume that all measurements are now done with a cadence $\Delta t_b = T/(n_{\rm vis} - 1)$ and a single intensity error. Using this resolution, we plot the covariance between the variability amplitude and timescale in Fig.~\ref{fig:ellipse}. In practice, the AGN shown in Fig.~\ref{fig:ellipse} are not affected by these assumptions given that we assume that only AGN that are detected at high signal to noise are included in the analysis.  
\subsection{AGN Population}
Given a set of individual AGN measurements compromising an AGN population, we also infer the precision with which we can measure the variability-timescale relation in Eq~\eqref{eq:tLrel}. Thus, we again carry out an expected information analysis using the power spectrum, but now parametrized by population parameters $A_b, \bar{t}_{b}^r,$ and $\beta_b$ and present the results in Fig.~\ref{fig:ellipse2}. Since we assume each AGN in this population is described by the same population parameters, the expected information is now the integral over the expected information gained from each of these AGN.  
\begin{comment} 
ow infer the covariance sensitivity $\sigma_{\beta_b}^2$ of an experiment to measure the timescale variability-luminosity index $\beta_b$. More specifically, we evaluate the ability of VRO to discern a population of AGN with variability parameters $\mathcal{V} = \{A, \bar{t}, \beta\}$ from a population undergoing no variability using the the expected information matrix from the power spectrum. 

We plot the covariance $({\rm SNR})_{\beta_b} = \left(\beta/\sigma_{\beta}\right)^2$ inferred from this matrix in Fig.~\ref{fig:betaSNR}. That is, we numerically invert the $3\times 3$ expected information matrix and take the $(3, 3)$ component. We find that for reasonable parameter values, the index $\beta_b$ will be able to be detected with high fidelity.
\end{comment}
\section{Discussion}\label{sec:disc}
Five assumptions are worth clarifying. First, we assumed that the AGN variability correlation function between two temporal measurements at $t_1$ and $t_2$ is only a function of the time lag $t = |t_2 - t_1|$, i.e. variability is  a stationary process. While this is often the case, non-stationarity has been found to exist under certain circumstances. If non-stationarity is a property of a particular class of AGN, then statistics such as the structure function or Wigner function may be utilized instead of the correlation function. 

Second, we modeled the correlation function using a damped random walk model, which as we stated previously, is not accurate for all AGN classes. However, for any two parameter model the forecasts presented should be accurate to within orders of unity. Models that include a third parameter, such as a damped random walk with an additional break in the corresponding power spectrum between the white and red noise regimes, will only reduce the fidelity of measurements of the variability amplitude and timescale and are outside the scope of this paper.

Third, we assume that the relationship between an AGN and its observed variability timescale can be described by two parameters: its redshift and bolometric luminosity. In reality, we expect other AGN parameters, such as its color, to also play an important role in determining the timescale within a class of AGN. Such a description of an AGN's variability timescale, while important and necessary for a complete description, is outside the scope of this paper. 

Fourth, we assumed that the observed frequency of light in a given band is the result of emitted light in the same frequency band. In reality, it is possible that light emitted in a higher frequency band will redshift across lower bands - leading to the final signal be a sum over different frequency bands. As a result, the autocorrelation of a single observed band will be the result of a cross correlation of emitted bands. Moreover, while we focused on variability two-point functions within a given band, the cross correlation between bands of VRO, as well as between VRO and other experiments will yield even more information about the structure of the AGN. Time lag measurements between UV/optical light and X-rays have already been used to measure the size regions such as the dust torus and broad-line region. We leave all such calculations for future work.

Lastly, we assumed that the true power spectrum can be recovered through measurements of the power spectrum in a finite box with finite resolution perfectly. For an actual experiment, we expect that measurements of the true power spectrum at Fourier modes close to either limit to be degraded. This degradement can be added in our expected information analysis through the introduction of an additional source of error. However, such error only has an effect on our final result when the variability timescale of an AGN becomes close to either limit. For a population of AGN, a bulk of them will most likely have variability timescales greater than a few days and less than a few years. As a result, we expect such degrading to not have a drastic impact on our results. 
\section{Conclusion}\label{sec:conc}
In this paper we presented a general framework for measuring the variability amplitude and timescale of any AGN. First, we measured the variability for each AGN and from there construct estimators for the variability correlation function. 

Since each timescale estimator was created using the power spectrum at two distinct modes, this introduced covariance between each timescale estimator. However, despite this covariance matrix being non-diagonal, we were able to calculate its inverse. Then, with each timescale estimator and the corresponding covariance matrix, we created a single estimator for the variability timescale using inverse covariance weighting. With an estimator for the variability amplitude and timescale, we then used linear error propagation to calculate the covariance matrix between these two parameters from the initial variability two-point functions.   
Using this covariance matrix, we forecasted the sensitivity of a VRO to measuring these parameters. We found that both the variability amplitude and timescale will be able to be measured up to $10\sigma$ across all bands. 

Finally, we calculated the theoretical best sensitivity to a VRO-like experiment measuring a relationship between the luminosity of an AGN and its variability amplitude and timescale. Namely, we used a logarithmic power law model between the luminosity of the AGN and its variability parameters. We found its index to be measured with at least $10^4\sigma$ fidelity.
 
\subsection*{Acknowledgments}
CCS acknowledges the support of the Bill and Melinda Gates Foundation. CCS was supported by a National Science Foundation Graduate Research Fellowship under Grant No.\ DGE-1746891.  This work was supported in part by the Simons Foundation and by National Science Foundation grant No.\ 2112699.

\end{document}